\documentclass[12pt]{article}
\usepackage{times}
\usepackage{psfig}

\newenvironment{sciabstract}{%
\begin{quote} \bf}
{\end{quote}}

\newcounter{lastnote}

\title{Quantum State Transfer Between Matter and Light}

\author
{D.~ N.~ Matsukevich, A.~Kuzmich $^{\ast}$ \\
\\
\normalsize{School of Physics, Georgia Institute of Technology,}\\
\normalsize{Atlanta, GA 30332, USA}\\
\\
\normalsize{$^\ast$To whom correspondence should be addressed;
E-mail:  alex.kuzmich@physics.gatech.edu.} }

\date{}

\begin{document}


\baselineskip24pt


\maketitle


\begin{sciabstract}
We report on the coherent quantum state transfer from a two-level
atomic system to a single photon. Entanglement between a single
photon (signal) and a two-component ensemble of cold rubidium
atoms is used to project the quantum memory element (the atomic
ensemble) onto any desired state  by measuring the signal in a
suitable basis. The atomic qubit is read out by stimulating
directional emission of a single photon (idler) from the
(entangled) collective state of the ensemble. Faithful atomic
memory preparation and read-out are verified by the observed
correlations between the signal and the idler photons. These
results enable implementation of distributed quantum networking.
\end{sciabstract}

\section*{Introduction}

The ability to coherently transfer quantum information between
photonic- and material-based quantum systems is a prerequisite for
all practical distributed quantum computation and scalable quantum
communication protocols ({\it 1}).  The importance of this process
is rooted in the fact that matter-based quantum systems provide
excellent long-term quantum memory storage, whereas long-distance
communication of quantum information will most certainly be
accomplished by coherent propagation of light, often in the form
of single photon pulses.

In the microwave domain,  coherent quantum control has been
obtained with single Rydberg atoms and single photons ({\it 2});
significant advances have also been made in ion trapping
information processing ({\it 3-5}). Particularly, an entangled
state of an ion and a photon has been produced ({\it 6}); however,
to convert a single ion (atom) qubit state into a photonic state,
strong coupling to a single cavity mode is required. Trapped atoms
or ions localized inside of high-finesse cavities offer a natural
paradigm for coherent, reversible matter-light interactions ({\it
7,8}), although technical challenges make these systems difficult
to realize in practice.

 Optically thick atomic ensembles have emerged recently as an alternative
 for the light-matter interface ({\it 9,10}). Duan, Lukin,
 Cirac and Zoller (DLCZ) ({\it 11}) have made a theoretical
 proposal aimed at long-distance quantum communication that utilizes
the quantum memory capability  of atomic ensembles. Important
initial steps towards realization of the DLCZ protocol have been
made in which non-classical radiation has been produced from an
atomic ensemble, thereby demonstrating the collective enhancement
({\it 12-15}).

Here we report on the experimental realization of coherent quantum
state transfer from a matter qubit onto a photonic qubit,
utilizing an optically thick cold atomic cloud. Our experiment
involves three steps: 1) an entangled state between a single
photon (signal) and a single collective excitation distributed
over many atoms in two distinct optically thick atomic samples is
generated. 2) measurement of the signal photon projects the atomic
ensembles into a desired state, conditioned on the choice of the
basis and the outcome of the measurement. This atomic state is a
nearly maximally entangled state between two distinct atomic
ensembles. 3) this nearly maximally entangled atomic state is
converted into a single photon (idler) emitted into a well-defined
mode, without using a high-finesse cavity. These three ingredients
constitute a complete set of tools required to build an arbitrary
large-scale quantum network ({\it 11}).

As illustrated in Fig.1A, the classical laser pulses used in the
generation and verification procedures define the two distinct
pencil-shape components of the atomic ensemble that form our
memory qubit, L and R. Fig.1B indicates schematically the
structure of the four atomic levels involved, $|a\rangle
,|b\rangle , |c\rangle $ and $|d\rangle $. The experimental
sequence starts with all of the atoms  prepared in state
$|a\rangle $. A write pulse tuned to the ${|a\rangle \rightarrow
|c\rangle }$ transition is split into two beams by a polarizing
beam splitter (PBS1) and passed through the atomic sample. The
light induces spontaneous Raman scattering on the ${|c\rangle
\rightarrow |b\rangle }$ transition. The classical write pulse is
so weak that less than one photon is scattered in this manner into
the forward direction mode for each pulse in either L or R. The
forward scattered mode is dominantly correlated with a distinct
collective atomic state ({\it 11}). In the first order of
perturbation theory in the atom-light coupling $\chi$, the
atom-light state is
\begin{eqnarray}
|\Psi \rangle \sim |a\rangle _{1} \ldots |a\rangle
_{N_L+N_R}|0_p\rangle _L|0_p\rangle _R +\chi (|L_a\rangle
|1_p\rangle _L|0_p\rangle _R+ |R_a\rangle |0_p\rangle _L|1_p
\rangle _R). \label{state}
\end{eqnarray}
We have defined two effective states of the atomic ensembles:
\begin{eqnarray}
|L_a\rangle =\sum_{i=1}^{N_L} g_i |a\rangle _{1}\ldots  |b\rangle
_{i}\ldots |a\rangle _{N_L} \ldots |a\rangle _{N_L+N_R} \nonumber \\
 |R_a\rangle =\sum_{j=N_L+1}^{N_L+N_R} g_j |a\rangle _{1}\ldots |a\rangle
_{N_L}\ldots |b\rangle _{j}\ldots |a\rangle _{N_L+N_R} ,
\end{eqnarray}
with the weights $g_i, g_j$ determined by the write field
intensity distribution, $\sum_{i=1}^{N_L} |g_i|^2 =1 $,
$\sum_{j=N_L+1}^{N_L+N_R} |g_j|^2 =1 $ ({\it 16,17}).
$|L_a\rangle$ and $|R_a\rangle$ have properties of a two-level
system (qubit): $\langle L_a|L_a\rangle =1$, $\langle
R_a|R_a\rangle =1$, $\langle L_a|R_a\rangle =0$. Although the
interaction of the light with the atoms is non-symmetric with
respect to permutation of atoms, the second term in Eq.\ref{state}
in fact describes a strongly entangled atom-photon state in the
sense of ({\it 17}). Using PBS4 and a half wave plate inserted
into one of the channels, we map the two spatial modes associated
with the two ensembles into a single spatial mode with
polarization encoding of the light's origin: $|1_p\rangle _L
\rightarrow |H\rangle _s; |1_p\rangle _R \rightarrow |V\rangle
_s$, where $H$ and $V$ indicate horizontal and vertical
polarization, respectively, and $s$ denotes signal. Next, the
light is passed through an arbitrary polarization state
transformer $R_s(\theta _s, \phi _s)$ and a polarizer PBS5, so
that the state at the output of PBS5 is
$$
|H^{\prime }\rangle = \cos ( \theta _s) e^{ i\phi _s} |H\rangle
_s+\sin (\theta _s) |V\rangle _s,
$$
and is directed onto a single-photon detector D1. When D1 detects
a photon, the joint state in Eq. 1 is projected into the desired
atomic state
\begin{equation}
|\Psi _a \rangle = \cos ( \theta _s) e^{ -i\phi _s} |L_a\rangle +
\sin ( \theta _s) e^{i\eta _s} |R_a\rangle,
\end{equation}
which is an entangled state of the two atomic samples $L$ and $R$.
Phase $\eta _s$ is determined by the difference in length of the
two paths $L$ and $R$. After a variable delay time $\Delta t$ we
convert the atomic excitation into a single photon by illuminating
the atomic ensemble with a pulse of light near resonant with the
$|b\rangle \rightarrow |d\rangle $ transition. For an optically
thick atomic sample, the photon will be emitted with high
probability into the spatial mode determined by the write pulse
({\it 11,16}), achieving memory read-out:
\begin{equation}
|\Psi _a \rangle  = \cos ( \theta _s ) e^{ -i\phi _s}|L_a \rangle
+ \sin (\theta _s) e^{i\eta _s}|R_a\rangle  \rightarrow |\Psi
\rangle _i = \cos ( \theta _s) e^{ -i\phi _s}|H\rangle _i  + \sin
(\theta _s ) e^{i(\eta _i +\eta _s) }|V\rangle _i .
\end{equation}
That is, the polarization state of the idler photon $i$ is
uniquely determined by the observed state of the signal photon.
Alternatively, one could  store the signal in a fiber until after
the read-out. In that case, the two-photon signal-idler state
would be a maximally entangled state:
\begin{equation}
|\Psi _M\rangle =\frac{1}{\sqrt{2}}(|H\rangle _s|H\rangle _i +
e^{i(\eta _s + \eta _i) }|V \rangle _s |V\rangle _i). \label{psim}
\end{equation}

A magneto-optical trap (MOT) of $^{85}$Rb is used to provide an
optically thick atomic cloud for our experiment (Fig.1).  The
ground states $\{|a\rangle;|b\rangle \}$ correspond to the
$5S_{1/2},F=\{3,2\}$ levels of $^{85}$Rb, while the excited states
$\{|c\rangle;|d\rangle \}$ represent the $\{5P_{3/2},F=3;
5P_{1/2},F=2\}$ levels of the $\{D_2, D_1\}$ lines at
$\{780;795\}$ nm, respectively. The experimental sequence starts
with all of the atoms  prepared in state $|a\rangle $ via optical
pumping, after shutting off the trapping and cooling light.

A 140 ns long  write pulse tuned to the ${|a\rangle \rightarrow
|c\rangle }$ transition is split into two beams by a polarizing
beamsplitter PBS1 and focused into two regions of the MOT about 1
mm apart with Gaussian waists of about $50$ $\mu m$. PBS2 and PBS3
separate the horizontally polarized component of the forward
scattered light from the vertically polarized classical pulse.
After being mixed by PBS4, the light goes through the quarter- and
the half-wave plates that provide the state transformation
$R_s(\theta _s ,\phi _s)$. The light continues to another
polarizer PBS5, and is directed to a single photon detector D1.
Detection of one photon by D1 prepares the atomic ensemble in any
desired state in the basis of $|L_a\rangle,|R_a\rangle $
determined by $R_s (\theta _s,\phi _s )$ and thereby concludes the
preparation of the quantum memory qubit.

Following memory state preparation, the read-out stage is
performed. After a user-programmable delay $\Delta t$, a 115 ns
long read pulse tuned to the ${|b\rangle \rightarrow |d\rangle }$
transition illuminates the two atomic ensembles. This accomplishes
a transfer of the memory state onto the single photon (idler)
emitted by the ${|d\rangle \rightarrow |a\rangle }$ transition.
After passing through the state transformer $R_i (\theta _i ,\phi
_i)$ and PBS6, the two polarization components are directed onto
single-photon detectors (D2, D3) thus accomplishing measurement of
the idler photon, and hence the memory qubit, in a controllable
arbitrary basis.

As in any real experiment, various imperfections  prevent
 the read-out of the quantum memory (idler photon) from being
 identical to the state that we intended to write into the
memory. To quantify the degree to which we faithfully prepare and
read-out the quantum memory, we measure the polarization
correlations between the signal and idler photons. The observed
correlations allow us to characterize the extent to which our
procedures are working. To investigate the storage capabilities of
our memory qubit quantitatively, we use time-resolved detection of
the signal and idler photons for two values of delay $\Delta t$
between the application of the write and read pulses, 100 ns and
200 ns. The electronic pulses from the detectors are gated with
250 ns and 140 ns windows centered on the time determined by the
write and read light pulses, respectively. Afterwards, the
electronic pulses are fed into a time-interval analyzer (with
$\delta =2$ ns time resolution). In order to measure the
correlation between the photons produced by the write and read
pulses, the output of D1 is fed into the ``Start" input of a
time-interval analyzer, and the outputs of D2 and D3 are fed into
two ``Stop" inputs. A coincidence window imposed by the data
acquisition software selects a time interval between the arrival
of the idler and signal of $(0,80)$ ns for $\Delta t=100$ ns and
$(25,145)$ ns for $\Delta t=200$ ns.

We first measure the conditional probabilities of detecting a
certain state of the idler (hence, of the quantum memory state) in
the basis of $|H\rangle _i$ and $|V\rangle _i$, given the observed
state of the signal photon. Varying the angle $\theta _s$ produces
the correlation patterns shown in Fig.2A for $\Delta t = 100$ ns.
Conditional probabilities at the point of maximum correlation are
shown in Fig.2B and the first line of Table 1. To verify faithful
memory preparation and read-out, we repeat the correlation
measurement in a different basis, of states $(|H\rangle _i \pm |V
\rangle _i)/ \sqrt{2}$, by choosing the $\theta _i =45$ degrees,
$\phi _i =0$ degrees, and $\phi _s=-(\eta _s + \eta _i )$ in the
state transformers $R_s$ and $R_i$. We vary $\theta _s$, with the
measured interference fringes displayed in Fig. 3A. Table 1
(second line) and Fig.   3B show the conditional probabilities at
the point of maximum correlations. These probabilities are
different from $1/2$ only when the phase coherence between the two
states of the atomic qubit is preserved in the matter-to-light
quantum state mapping.

From these measured correlations, we determine the fidelity of the
reconstruction of our intended
 quantum memory state $| \Psi_{I}\rangle $
in the idler, $|\langle \Psi_{I}|\Psi _{i} \rangle
 |^2$. The fidelity is given by the value of the corresponding
 conditional probability at the point of maximum correlation, presented in
 Table 1 (we choose the lower of the two values as the lower bound).
 For states in the $\theta _i =0$ degree basis, we find $F_0 =0.88 \pm
 0.03$, clearly exceeding the classical boundary of $2/3$ ({\it 18}). For the
 $\theta _i =45$ degree basis, we found $F_{45}=0.75\pm 0.02$, again
 significantly violating the classical limit. These fidelities  give a
lower bound for both the fidelities  of the memory preparation and
the read-out steps, which we do not measure separately.

Another way to quantify the performance of our quantum state
transfer is to calculate the fidelity of entanglement between the
signal and idler photons $F_{si}$. The lower bound on $F_{si}$  is
given by the overlap of the measured density matrix with the
maximally entangled state we seek to achieve $|\Psi _M\rangle $
given by Eq.\ref{psim}: $ F_{si}= \langle \Psi _M | \rho _{si} |
\Psi _{M}\rangle $ ({\it 19}). We calculated $F_{si} =0.67 \pm
0.02$, substantially greater that the classical limit of $1/2$
({\it 6,19}).

At a longer delay of 200 ns the fidelities in the $\theta _i =0$
degrees and $\theta _i =45$ degrees bases are $F_0= 0.79 \pm 0.04$
and $F_{45}=0.74 \pm0.04$, while fidelity of entanglement is
$F_{si} =0.63 \pm 0.03$.  For both values of $\Delta t$, we
analyze the fidelity of entanglement as a function of the delay
between the detections of the signal and the idler. We split the
full coincidence window into four equal intervals, and calculated
entanglement of formation for each one (Fig.4). From these
results, we conclude that our quantum memory has a useful
operational time of about 150 ns. The lifetime of coherence
between the levels $|a\rangle $ and $|b\rangle $ determines the
lifetime of the quantum memory and is limited by the magnetic
field of the trapping quadrupole field of the MOT ({\it 12}).

Non-zero coincidence counts in the minima of Fig. 2A are due to
transmission losses and non-ideal spatial correlations between the
signal and idler photons. The residual interferometric drifts in
$\eta _s + \eta _i$ further reduce the visibility of Fig. 3A
compared to Fig. 2A, resulting in a degradation of the fidelities.
Losses also reduce the rate of entanglement generation. The rate
of signal photon detections (and hence, atomic qubit preparation)
is given by $R_s = \alpha n_s R \simeq 300 s^{-1}$, where $\alpha
= 0.05$ is the measured transmission efficiency for the write beam
(which includes $0.60$ detection efficiency), and $R = 4.7 \times
10^5 s^{-1}$ is the repetition rate of the experiment. Therefore,
the inferred average photon number in the forward scattered mode
per pulse is $n_s \simeq 1.4\times 10^{-2}$. The coincident
signal-idler detection rate is $R_{si} = \zeta R_s =\zeta \alpha
{n_s} R \simeq 0.4 s^{-1}$, where  $\zeta \equiv \beta \xi \simeq
1.1 \times 10^{-3}$. The measured transmission and detection
efficiency for the read beam is $\beta \simeq 0.04$, so we infer
the efficiency of quantum state transfer from the atoms onto the
photon $\xi \simeq 0.03$.

We have realized a quantum node by combining the entanglement of
an atomic and photonic qubits with the atom-photon quantum state
transfer. By implementing the second node at a different location,
and performing a joint detection of the signal photons from the
two nodes, the quantum repeater protocol ({\it 11}), as well as
distant teleportation of an atomic qubit may be realized. Based on
this work, we estimate the rate for these protocols to be
$R_{2}\simeq (\beta \xi \alpha n_s)^2 R \simeq 3\times 10^{-7}
s^{-1}$. However, improvements in $\xi $ that are based on
increasing the optical thickness of atomic samples ({\it 16}), as
well as elimination of transmission losses could provide several
orders of magnitude increase in $R_2$. Our results also
demonstrate the possibility of realizing quantum nodes consisting
of multiple atomic qubits by using multiple beams of light. This
approach shows promise for implementation of distributed quantum
computation ({\it 20,21}).

\begin{quote}
{\bf References and Notes}

\begin{enumerate}
\item I. Chuang, M. Nielsen, {\it Quantum computation and quantum
information}, (Cambridge University Press, 2000).
\item S. Haroche, J. M. Raimond, M. Brune, in {\it Experimental Quantum
Computation and Information}, eds. F. de Martini and C. Monroe),
37-66 (Proc. Int. School of Physics Enrico Fermi, course CXLVIII,
IOS Press, Amsterdam, 2002).
\item C. A. Sackett et al.,  {\it Nature} {\bf 404},
256 (2000).
\item M. D. Barrett et al, {\it Nature} {\bf 429},
737 (2004).
\item M. Riebe et al,  {\it Nature} {\bf 429},
734 (2004).
\item B. B. Blinov, D. L. Moehring, L.-M. Duan, C. Monroe, {\it
Nature} {\bf 428}, 153 (2004).
\item S. Bose, P. L. Knight, M. B. Plenio,
V. Vedral,  {\it Phys. Rev. Lett.} {\bf 83}, 5158 (1999).
\item H. J. Kimble, Phys. Scr. {\bf 76}, 127 (1998).
\item A. Kuzmich, E. S. Polzik, in {\it Quantum
information with continuous variables}, (eds. S. L. Braunstein and
A. K. Pati, Kluwer, 2003).
\item  M. D. Lukin,  {\it Rev. Mod. Phys.} {\bf 75}, 457 (2003).
\item L.-M. Duan, M. D. Lukin, I. J. Cirac, P. Zoller, {\it
Nature}
{\bf 414}, 413 (2001).
\item A. Kuzmich et al., {\it Nature} {\bf 423}, 731
(2003).
\item C. H. van der Wal et al., {\it Science} {\bf 301}, 196 (2003).
\item W. Jiang, C. Han, P. Xue, L.-M. Duan, G. C. Guo, {\it Phys. Rev.
A} {\bf 69}, 043819 (2004).
\item C. W. Chou, S. V. Polyakov, A. Kuzmich, H. J. Kimble, {\it Phys.
Rev. Lett.} {\bf 92}, 213601 (2004).
\item L.-M. Duan, J. I. Cirac,  P. Zoller, Phys. Rev. A
{\bf 66}, 023818 (2002).
\item A. Kuzmich, T. A. B. Kennedy, {\it  Phys. Rev. Lett.} {\bf 92}, 030407 (2004).
\item M. Horodecki, P. Horodecki, R. Horodecki, {\it Phys. Rev. A} {\bf 60}, 1888 (1994).
\item  C. H. Bennett, D. P. DiVincenzo, J. A. Smolin, W. K.
Wooters, Phys. Rev. A {\bf 54}, 3824 (1996).
\item Y.L. Lim, A. Beige, L.C. Kwek, www.arXiv.org/quant-ph/0408043.
\item S. D. Barrett, P. Kok, www.arXiv.org/quant-ph/0408040.
\item We acknowledge fruitful conversations with T. A. B. Kennedy,
J. A. Sauer, L. You, A. Zangwill and, particularly, M. S. Chapman,
and thank R. Smith and E. T. Neumann for experimental assistance.
This work was supported by NASA and the Research Corporation.
\end{enumerate}
\end{quote}

\clearpage

\begin{table}
\caption{\label{tab:table1} Conditional probabilities $P(I|S)$ to
detect the idler photon in state $I$ given detection of the signal
photon in state $S$, at the point of maximum correlation for
$\Delta t = 100$ ns delay between read and write pulses; all the
errors are statistical.}
\vspace{1.0 cm} \centerline{
\begin{tabular}{ccccc}
Basis  & $P(H_i|H_s)$ & $P(V_i|H_s)$  & $P(V_i|V_s)$ & $P(H_i|V_s)$ \\
\hline
$0$    & $0.92 \pm 0.02$  & $0.08 \pm 0.02$ & $0.88 \pm 0.03$ & $0.12 \pm 0.03$ \\
$45$   & $0.75 \pm 0.02$  & $0.25 \pm 0.02$ & $0.81 \pm 0.02$ & $0.19 \pm 0.02$ \\
\end{tabular}
}
\end{table}

\clearpage

\begin{figure}
\begin{center}
\leavevmode  \psfig{file=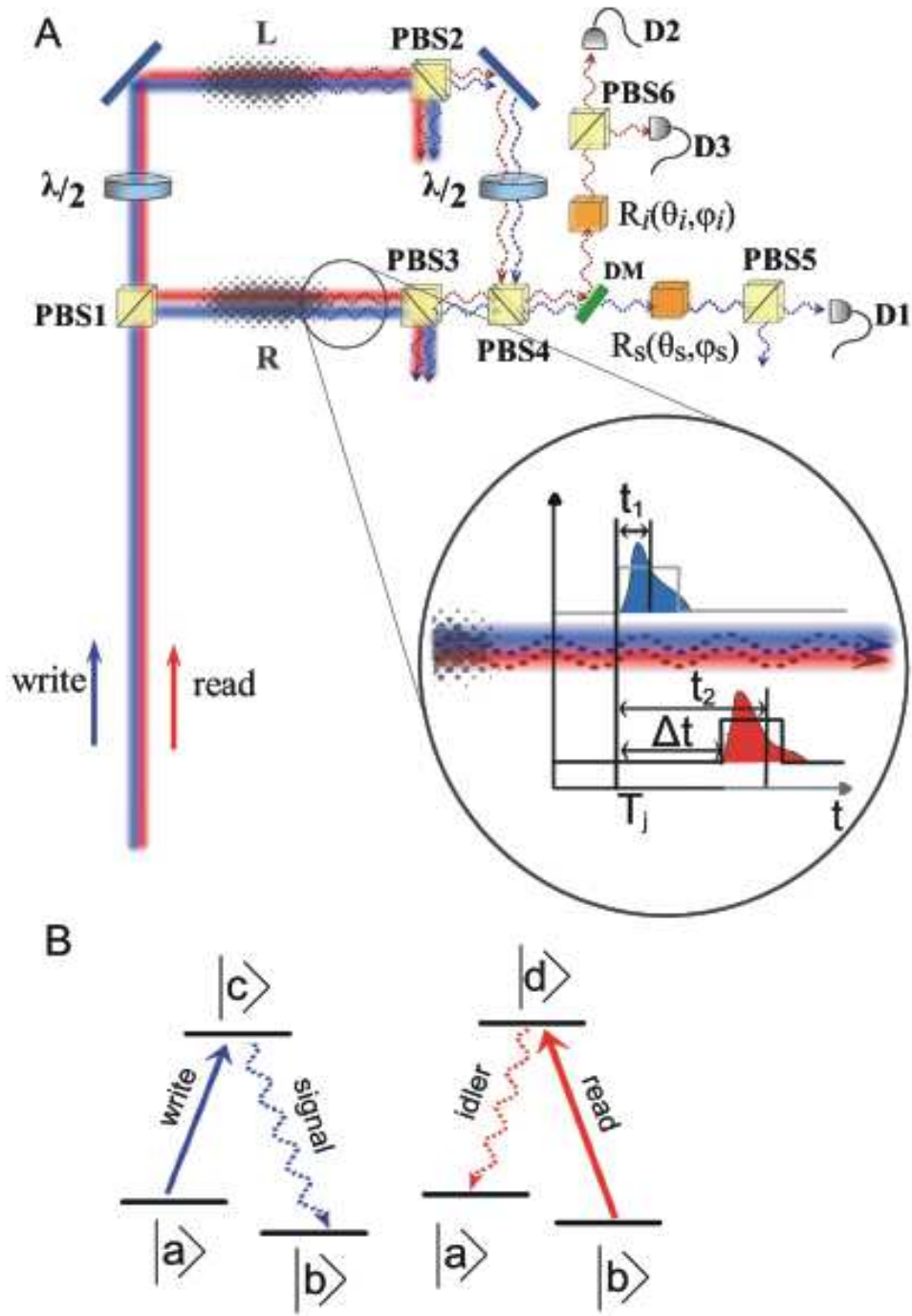,width=5in}
\end{center}
\caption{ (A)   Schematic of experimental setup. PBS1-6,
polarizing beam splitters, $\lambda /2$, half waveplate,
polarization state transformers, $R_s(\theta _s, \phi _s)$ and
$R_i(\theta _i, \phi _i)$, (D1,D2,D3), single photon detectors,
DM, dichroic mirror. The inset illustrates the timing of the write
and read pulses. (B) The relevant atomic level structure.}
\end{figure}

\begin{figure}
\begin{center}
\leavevmode  \psfig{file=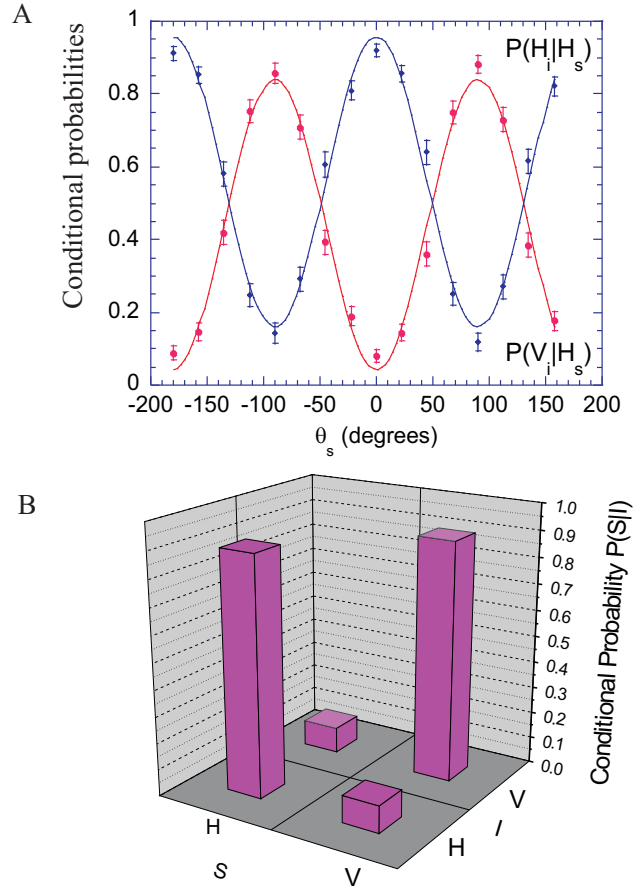,height=5in}
\end{center}
\caption{ (A) Measured conditional probabilities $P(H_i|H_s)$ and
$P(V_i|H_s)$ as the function of the polarization rotation $\theta
_s$ of the signal photon. The full curves are fits with the
visibility as the only adjustable parameter. (B) Measured
conditional probabilities at the points of highest correlation.}
\end{figure}

\begin{figure}
\begin{center}
\leavevmode  \psfig{file=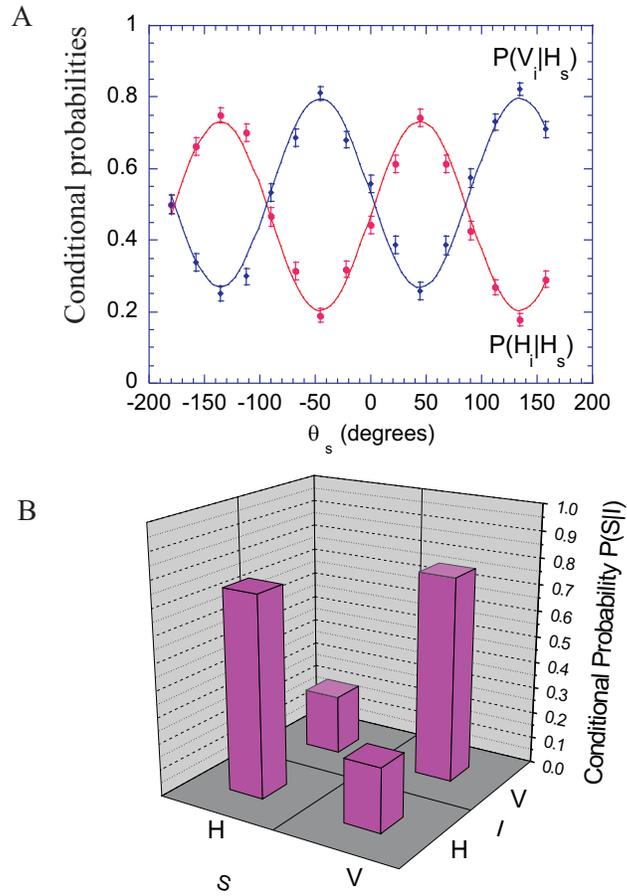,height=5in}
\end{center}
\caption{ (A) Measured conditional probabilities after $\theta _i
=\pi/4$ polarization rotation  of the idler photon as the function
of $\theta _s$. (B) Measured conditional probabilities at the
points of highest correlation.}
\end{figure}

\begin{figure}
\begin{center}
\leavevmode  \psfig{file=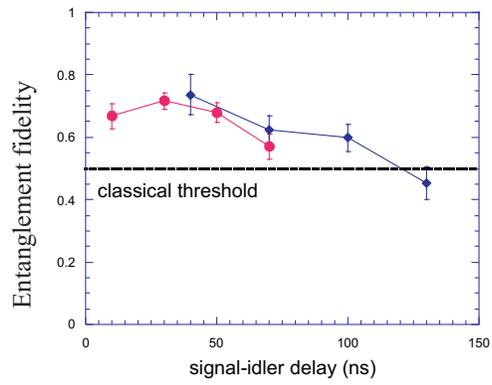,width=2.5in}
\end{center}
\caption{Time-dependent entanglement fidelity of the signal and
the idler $F_{si}$; circles for $\Delta t=100$ ns, diamonds for
$\Delta t=200$ ns.}
\end{figure}

\end{document}